\documentclass[twocolumn,10pt]{asme2ej}

\usepackage{graphicx} 
\usepackage{amsmath, amssymb}
\usepackage{bm}
\usepackage{hyperref}
\usepackage{color}
\usepackage{url}

\usepackage[usernames,dvipsnames]{xcolor}

\title{Dynamic modeling of a sliding ring on an elastic rod with incremental potential formulation}

\author{Weicheng Huang \thanks{Corresponding author.}
    \affiliation{
		School of Engineering, \\ Newcastle University, \\ Newcastle upon Tyne NE1 7RU, UK \\
        Email: weicheng.huang@newcastle.ac.uk
    }	
}

\author{Peifei Xu \thanks{Corresponding author.}
    \affiliation{
    School of Mechanical Engineering, \\ Southeast University, \\ Nanjing, 211189, China. \\
    Email: xupeifei@seu.edu.cn
    }   
}

\author{Zhaowei Liu \thanks{Corresponding author.} 
    \affiliation{
    College of Mechanics and Materials, Hohai University, \\ Nanjing 211100, People's Republic of China \\
        Email: zhaowei.liu@hhu.edu.cn
    }   
}

\begin{document}

\maketitle    

\begin{abstract}
{\it 
Mechanical interactions between rigid rings and flexible cables find broad application in both daily life (hanging clothes) and engineering system (closing a tether-net).
A reduced-order method for the dynamic analysis of sliding rings on a deformable one-dimensional (1D) rod-like object is proposed.
In contrast to the conventional approach of discretising joint rings into multiple nodes and edges for contact detection and numerical simulation, a single point is used to reduce the order of the model.
To ensure that the sliding ring and flexible rod do not deviate from their desired positions, a new barrier function is formulated using the incremental potential theory. Subsequently, the interaction between tangent frictional forces is obtained through a delayed dissipative approach.
The proposed barrier functional and the associated frictional functional are $C^{2}$ continuous, hence the nonlinear elastodynamic system can be solved variationally by an implicit time-stepping scheme.
The numerical framework is initially applied to simple examples where the analytical solutions are available for validation. Then, multiple complex practical engineering examples are considered to showcase the effectiveness of the proposed method.
The simplified ring-to-rod interaction model has the capacity to enhance the realism of visual effects in image animations, while simultaneously facilitating the optimisation of designs for space debris removal systems.
}
\end{abstract}

\section{Introduction}
\label{sec:Intro}

Sliding rings on a flexible rod is a commonly used structure in daily life, such as clotheslines and freight transport. The relevant research can be traced back to the last century~\cite{mote1966nonlinear,tabarrok1974dynamics,wickert1992non}.
Moreover, due to its high efficiency for structural connections, the ring-rod mechanism is also widely used in practical engineering, including cable-driven system~\cite{zheng2022ale}, deployable structures~\cite{tonoli2014modeling,fu2019multiscale}, drive belts~\cite{tonoli2014modeling,vetyukov2019flexible}, and continuum manipulators/robots~\cite{ma2020dynamic,xu2018kinematics}.
Especially, the space tether-net system~\cite{lim2018dynamic,chen2014dynamical}, a novel lightweight and cost-efficient technique for space debris removal~\cite{botta2016simulation,botta2019simulation,shan2019contact,hou2021dynamic}, usually employs several joint rings to connect tether rods and flexible nets, in order to achieve the closing mechanism in the post-capture phase~\cite{sharf2017experiments,botta2020simulation,huang2022Nonlinear,huang2023contact,huang2023numerical}.
The central aspect in engineering structure design, spanning various applications such as bridges, railways, cable cars, and tether-nets, revolves around the development of a mathematical model capable of effectively characterising the nonlinear dynamic interaction between a flexible medium and a mobile ring.
On the one side, plenty of investigations have been conducted on the mechanical modeling of rigid particles, flexible rods, and their interactions.
The dynamics of a single point mass is straightforward, i.e., its equation of motion can be given directly by using Newtonian mechanics;
while the governing equations for 1D rods would be slightly more complicated, as the structural slenderness may cause geometrically nonlinear deformation.
To reveal the nonlinear mechanics of slender structures when experiencing finite deflections and rotations, tremendous numerical algorithms have been developed for engineering usages over the past few decades, such as Finite Element Method (FEM)~\cite{hughes2012finite}, Geometrically Exact Beam Formulation (GEBF)~\cite{sonneville2014geometrically,farokhi2022experimentally}, Absolute Nodal Coordinate Formulation (ANCF)~\cite{shabana1996absolute,shabana2001three,yakoub2001three}, and Isogeometric Collocation Methods (ICM)~\cite{kiendl2015isogeometric}.
Moving forward, even not exactly identical, the interplay between rigid rings and flexible rods can be partially treated as a sliding beam problem (e.g., classical sliding spaghetti)~\cite{steinbrecher2017numerical,han2022configurational}, and the dynamic governing equations can be derived using Lagrangian equations with non-material volume~\cite{irschik2002equations,pesce2003application}, extended Hamilton's principle~\cite{humer2020general,boyer2022extended}, and Arbitrary Lagrangian-Eulerian (ALE) formulation~\cite{fotland2022numerical}.
Regarding the sliding connection between the joint ring and the flexible medium, the majority of prior research predominantly focused on scenarios where friction was not taken into account.
%
Munoz and Jelenic employed the master–slave approach to formulate the sliding joint, aiming to establish a contact condition devoid of friction while minimizing the number of coordinates involved, as detailed in their work~\cite{munoz2004sliding}.
%
Lee et al. developed a three-dimensional (3D) sliding joint for nonlinear beams concerning the discontinuity using finite element discretization~\cite{lee2008development}. 
An ANCF-ALE coupled framework was proposed by Hong et al. to analysis the frictionless interaction between sliding joint and flexible beam~\cite{hong2011modeling}.
Recently, isogeometric interpolation was used for frictionless sliding contact coupling dynamics to achieve higher smoothness and strong robustness~\cite{Guo2022Energy}.

On the other side, within the computer-aided geometric design (CAGD) community, visual animations for elastodynamic systems often make use of a different class of numerical frameworks known as Discrete Differential Geometry (DDG) methods. This choice is driven by their resilience and computational efficiency in simulating slender elastic structures undergoing geometrically nonlinear deformation, as well as handling frictional contact~\cite{huang2020dynamic} and fluid-structure interactions~\cite{jawed2015propulsion}. Examples of such applications include the simulation of phenomena like hair fluttering and cloth wrinkling.
The DDG method for numerical simulation begins by discretizing a continuous structure into a system of mass-spring-damper elements, all the while maintaining its essential geometric characteristics. This approach is employed to accurately represent and capture the nonlinear deformations of the structure~\cite{grinspun2006discrete}.
%
Previous DDG based numerical frameworks have shown surprising performance in simulating thin elastic structures, such as rods~\cite{bergou2008discrete,bergou2010discrete}, ribbons~\cite{shen2015geometrically,huang2021snap}, plates/shells~\cite{baraff1998large,huang2020shear}, and gridshells/nets~\cite{panetta2019x,huang2021numericalmethod}.
Furthermore, to gain acceptance within the mechanical, civil, and aerospace engineering communities, a substantial volume of practical tabletop experiments have recently been undertaken to validate the accuracy of these visually striking simulations. These experiments encompass a range of phenomena, such as the coiling of rods~\cite{jawed2014coiling}, the large deflection of strips~\cite{romero2021physical}, the bifurcation of plates~\cite{huang2020shear}, and the form-finding of gridshells~\cite{baek2018form}.
%
Regarding the contact mechanics within flexible multibody systems, the incremental potential method introduced in the year 2020 is widely considered to be one of the most effective frameworks~\cite{li2020incremental}.
%
The incremental potential formulation commences by introducing a barrier functional that maintains $C^{2}$ continuity. This functional is based on the Euclidean distance between two approaching particles (or other geometric elements like edges and surfaces, as discussed in~\cite{li2020codimensional}). Subsequently, the formulation proceeds to analytically derive the gradient vector, which corresponds to the repulsive force, and the Hessian matrix, which represents the tangential stiffness associated with this barrier functional.
%
The frictional force, similarly, is given by maximal dissipation principle with a velocity-related normalized parameter to increase functional continuous.
{The incremental potential model and its extension have already shown powerful performance when dealing with complex rod-to-rod interaction and soft robotic dynamics, e.g., knot tying~\cite{choi2021implicit}, flagella bundling~\cite{tong2022fully}, and magnetic cilia locomotion~\cite{huang2023modeling}.}
Nonetheless, in cases where both the ring and the rod are discretized into nodes and edges and simulated using a well-established rod-to-rod interaction model, the computational time remains substantial due to the increased number of degrees of freedom. Furthermore, when dealing with a small-sized ring, it becomes necessary to employ a denser mesh and shorter time steps to ensure accurate contact detection and maintain numerical stability.

This manuscript aims to develop a reduced-order method for analysing the nonlinear dynamic interaction between a sliding ring on a flexible cable.
Different from the conventional contact mechanics, which discretizes the ring into nodes and edges to achieve penetration detection and intersection-free dynamics, the ring in the present approach is treated as a single point for model order reduction.
The non-penetration condition between a sliding joint and an elastic rod is transformed into a non-deviation condition using a novel barrier-like energy functional.
It is worth noting that the barrier functional for ring-to-rod interaction is based on an incremental potential formulation with $C^2$ continuity, such that its gradient vector and Hessian matrix are avaliable and the flexible multibody system can be solved variationally through an implicit time-stepping method~\cite{li2020incremental}.
The tangential frictional interaction is implemented based on the maximal dissipation principle in a manner similar to the normal contact formulation~\cite{moreau2011unilateral}.
To validate the precision and correctness of our numerical implementation, we initially apply our discrete simulation to simple cases. Specifically, we focus on scenarios involving a sliding ring on a rigid bar, considering both straight and curved configurations. These cases are chosen because analytical solutions for them are readily available.
%
Then, the stiffness of 1D medium is decreased to address the nonlinear dynamic coupling between a sliding ring and a deformable rod, in which the geometrically nonlinear deformation is considered.
Subsequently, one proceed to illustrate two intricate examples, each with the potential for practical applications in both computer graphics (such as simulating hanging clothes) and aerospace engineering (like closing a tether-net). These demonstrations serve to highlight the applicability of our robust model in solving real-world engineering problems.
%
In comparison to the conventional penalty methods, such as the linear spring method~\cite{hou2021dynamic}, the Incremental Potential formulation presents several notable advantages~\cite{li2020incremental}: (i) It exhibits $C^2$ continuity, enabling the obtainment of its gradient and Hessian, thereby enhancing the robustness of the implicit simulator; (ii) The IP formulation is free from intersections and inversions, which is crucial for dynamic rendering, ensuring physically plausible outcomes; (iii) It offers tunability and can be readily adapted to different material stiffness and desired levels of accuracy. For instance, when prioritizing accuracy, a smaller clamped distance and reduced time step size can be employed. On the other hand, when aiming for computational efficiency, a slightly larger clamped distance parameter and increased time step size can be utilised.
{Traditional $C^2$ continuous barrier energy functional has shown surprisingly successful performance when dealing with the contact and collision, and our newly-introduced energy functional is not unique; we just change the well-established IP formulation from non-penetration contact to non-deviation slide, and holds all other advantages in the IP method.}

The remainder of this manuscript is organized as follows.
In Sec. $2$, the ring-to-rod contact model and the associated numerical framework are formulated in detail.
The proposed numerical framework is applied onto multiple examples in Sec. $3$.
Lastly, conclusive remarks and future research are discussed in Sec. $4$.

\section{Numerical method}
\label{sec:NumericalMethod}

Within this section, one presents a numerical model designed for simulating contact interactions between a sliding ring and a flexible rod. This model is established through a modified incremental potential method.
Firstly, we provide a concise overview of DDG-based models applied to rod and plate structures. 
Subsequently, we delve into a discussion of the classical incremental potential formulation employed in modeling contact between particles.
This framework is then expanded to encompass the reduced-order modelling of interactions between a ring and a rod.
Lastly, we present the dynamic equations of motion governing the behaviour of the elastodynamic system.

\begin{figure}[h!]
  \centering
  \includegraphics[width=1.0\columnwidth]{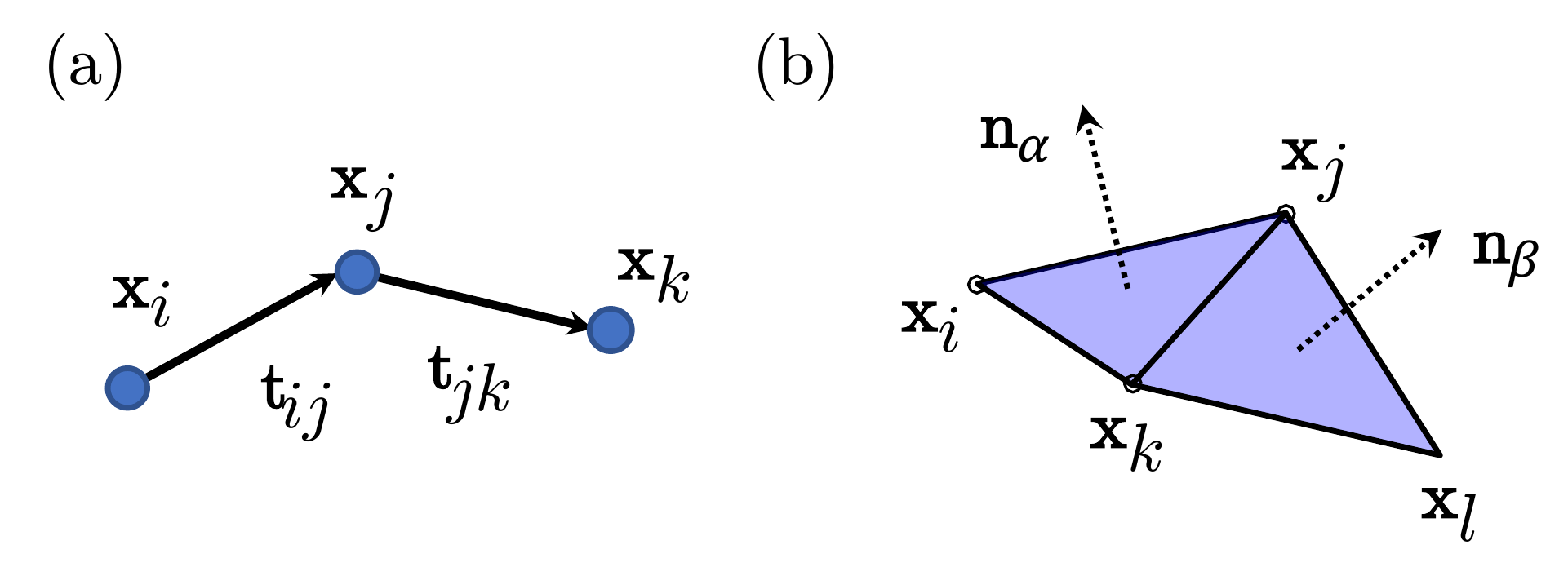}
  \caption{Discrete elements used in our numerical simulations. (a) Rod model. (b) Plate model.}
  \label{fig:discreteModelPlot}
\end{figure}

\subsection{Discrete models for thin elastic structures}
Here, one provides a brief overview of the DDG-based numerical frameworks applied to both 1D rod structures~\cite{bergou2008discrete,bergou2010discrete} and 2D plate objects~\cite{liang2009shape,savin2011growth}. These frameworks have undergone validation through comparisons with either experimental data or analytical solutions in previous studies~\cite{jawed2014coiling,huang2022Nonlinear,huang2020shear}.
%
The simulation domain is discretized into numerous nodes and elements, effectively conceptualized as a mass-spring-damper system. Within this system, a lumped mass is positioned at each vertex, accompanied by its corresponding discrete energy contributions.
%
Referring to Fig.~\ref{fig:discreteModelPlot}(a), the basic rod element combines $3$ vertices, $\mathcal{R}_{ijk} := \left(\mathbf{x}_{i},\mathbf{x}_{j},\mathbf{x}_{k} \right) \in \mathbb{R}^{9}$ with $2$ edges, which are defined by
\begin{subequations}
\begin{align}
\mathbf{e}_{ij} &= \mathbf{x}_{j} - \mathbf{x}_{i}, \\
\mathbf{e}_{jk} &= \mathbf{x}_{k} - \mathbf{x}_{j}, \\
\mathbf{t}_{ij} &= \frac {\mathbf{e}_{ij}}  {|\mathbf{e}_{ij}|}, \\
\mathbf{t}_{jk} &= \frac {\mathbf{e}_{jk}}  {|\mathbf{e}_{jk}|}, \\
\Delta l_{ijk} & = \frac {1}{2} (|\mathbf{e}_{ij} | + |\mathbf{e}_{jk}| ).
\end{align}
\end{subequations}
For a 1D rod with Young's modulus $E$, cross section area $A$, and moment of inertia $I$, the simplified discrete elastic energies are given by~\cite{huang2022Nonlinear}
\begin{subequations}
\begin{align}
E^{s}_{\mathcal{R}} &= \sum_{ij} \frac{1} {2}  {EA} \frac { \left({ | \mathbf{e}_{ij} |} - {| \bar{\mathbf{e}}_{ij} |} \right)^2 } {{| \bar{\mathbf{e}}_{ij} |} }, \\
E^{b}_{\mathcal{R}} &= \sum_{ijk} \frac{1}{2}  {EI} \frac { \left( | \mathbf{t}_{ij} - \mathbf{t}_{jk} | \right)^2} { {{\Delta \bar{l}_{ijk}}} },
\end{align}
\end{subequations}
where a bar on the top represents the quantity in reference configuration.
In this context, we have omitted consideration of the twisting energy associated with the rod's centerline. However, it is worth noting that if deemed necessary, this twisting energy can be readily incorporated by introducing an additional degree of freedom in the form of a twisting angle, as explained in prior works~\cite{bergou2008discrete,bergou2010discrete}.

The plate element is defined by $4$ vertices, $\mathcal{P}_{ijkl} := \left(\mathbf{x}_{i},\mathbf{x}_{j}, \mathbf{x}_{k}, \mathbf{x}_{l} \right) \in \mathbb{R}^{12}$ with $2$ triangular faces, which are given by
\begin{subequations}
\begin{align}
\mathbf{e}_{ij} &= \mathbf{x}_{j} - \mathbf{x}_{i}, \\
\mathbf{e}_{ik} &= \mathbf{x}_{k} - \mathbf{x}_{i}, \\
\mathbf{e}_{lj} &= \mathbf{x}_{j} - \mathbf{x}_{l}, \\
\mathbf{e}_{lk} &= \mathbf{x}_{k} - \mathbf{x}_{l}, \\
\mathbf{t}_{ij} &= \frac {\mathbf{e}_{ij}}  {|\mathbf{e}_{ij}|}, \\
\mathbf{t}_{ik} &= \frac {\mathbf{e}_{ik}}  {|\mathbf{e}_{ik}|}, \\
\mathbf{t}_{lj} &= \frac {\mathbf{e}_{lj}}  {|\mathbf{e}_{lj}|}, \\
\mathbf{t}_{lk} &= \frac {\mathbf{e}_{lk}}  {|\mathbf{e}_{lk}|}, \\
\mathbf{n}_{\alpha} &= \mathbf{t}_{ik} \times \mathbf{t}_{ij}, \\
\mathbf{n}_{\beta} &= \mathbf{t}_{lj} \times \mathbf{t}_{lk},
\end{align}
\end{subequations}
where $\mathbf{n}_{\alpha}$ and $\mathbf{n}_{\beta}$ are the surface normal vectors of the two triangular faces, referring to Fig.~\ref{fig:discreteModelPlot}(b).
For a 2D plate with Young's modulus $Y$ and thickness $b$, the discrete elastic energies are the sum of in-plane stretching and out-of-plane bending~\cite{savin2011growth},
\begin{subequations}
\begin{align}
E^{s}_{\mathcal{P}} &= \sum_{ij} \frac{\sqrt{3}} {4}  {Yb}  { \left({ | \mathbf{e}_{ij} |} - {| \bar{\mathbf{e}}_{ij} |} \right)^2 }, \\
E^{b}_{\mathcal{P}} &= \sum_{\alpha \beta} \frac{1}{12 \sqrt{3}}  {Yb^3} { \left( | \mathbf{n}_{\alpha} - \mathbf{n}_{\beta} | \right)^2}.
\end{align}
\end{subequations}

\subsection{Incremental potential method for contact}
The incremental potential contact method represents an effective numerical model for addressing nonlinear elastodynamic systems that involve contact~\cite{li2020incremental}.
%
Only particle-particle contact is considered in present method, and other arbitrary dimension (surfaces and curves) contact can be derived in a similar approach~\cite{li2020codimensional}.

\begin{figure*}[h!]
  \centering
  \includegraphics[width=1.0\textwidth]{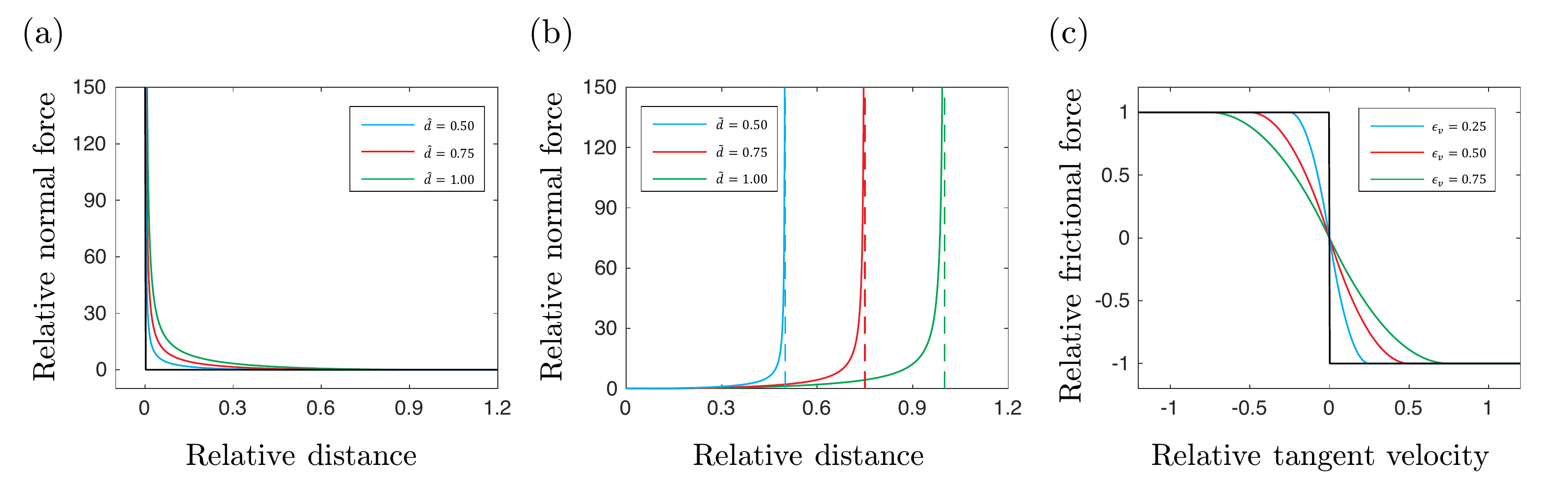}
  \caption{(a) Barrier potential for classical contact problem. (b) Modified barrier potential for the interaction between a ring and a rod. (c) Friction potential.}
  \label{fig:contactModelPlot}
\end{figure*}

When the Euclidean distance between two approaching nodes, $\mathcal{C}_{ij} := \left( \mathbf{x}_{i}, \mathbf{x}_{j} \right) \in \mathbb{R}^{6}$, is smaller than a threshold, the repulsive force needs to be included into the discrete dynamic system thus the non-penetration condition can be achieved.
The incremental potential model uses a nonlinear barrier-type potential with $C^{2}$ continuity to achieve the intersection free condition between two approaching nodes~\cite{li2020incremental},
\begin{equation}
E_{\mathcal{C}} = 
\begin{cases}
- K_{\mathcal{C}} \left[ (d - \hat{d})^2 \log (\frac{d} {\hat{d}}) \right] & \; \mathrm{ when } \; 0 < d < \hat{d} \\
0 & \; \mathrm{ when } \; d \geqslant \hat{d},
\end{cases}
\end{equation}
where $d = |\mathbf{x}_{j} -  \mathbf{x}_{i}|$ is the Euclidean distance between two considered nodes, $K_{\mathcal{C}}$ is the stiffness parameter, and $\hat{d}$ is a barrier parameter.
It is assumed here that the particle size is zero, and the relative distance can be easily derived by subtracting the particle radius if the ball with finite size is considered.
By taking the variation of nonlinear barrier potential, the contact force along the normal direction is given by~\cite{li2020incremental},
\begin{equation}
| \mathbf{F}_{\mathcal{C}}^{n} | = 
\begin{cases}
K_{\mathcal{C}} \left[ 2 (d - \hat{d}) \log (\frac{d} {\hat{d}}) + \frac {(d - \hat{d})^2}{d}  \right] & \; \mathrm{when} \; 0 < d < \hat{d} \\
0 & \; \mathrm{when} \; d \geqslant \hat{d}.
\end{cases}
\label{eq:barrierFunction}
\end{equation}
The contact force is zero when the distance is beyond the threshold, $\hat{d}$, and would gradually increase as the distance between two nodes is smaller.
The dependency of normalized contact force, $| \mathbf{F}_{\mathcal{C}}^{n} | / K_{\mathcal{C}}$, on the relative distance, $d$, for different barrier parameter, $\hat{d} \in \{0.50, 0.75, 1.00 \}$, is plotted in Fig.~\ref{fig:contactModelPlot}(a).

Then, the tangential component of contact force is applied through a dissipative potential form, which follows the maximal dissipation principle~\cite{moreau2011unilateral,li2020incremental},
\begin{equation}
| \mathbf{F}_{\mathcal{C}}^{t}| =
\begin{cases}
\mu | \mathbf{F}_{\mathcal{C}}^{n} | \left(- \frac {|\mathbf{v}^{t}|^2} {\epsilon_{v}^2} +  \frac {|\mathbf{v}^{t}|} {\epsilon_{v}} \right) & \; \mathrm{when} \; 0 \leqslant |\mathbf{v}^{t}| \leqslant \epsilon_{v}, \\
\mu | \mathbf{F}_{\mathcal{C}}^{n} | &  \; \mathrm{when} \;  |\mathbf{v}^{t}| > \epsilon_{v},
\end{cases}
\label{eq:barrierFriction}
\end{equation}
where $\mu$ is the frictional coefficient, $\mathbf{v}^{t}$ is the relative velocity along the tangential direction between two reference nodes, and $\epsilon_{v}$ is the velocity magnitude bound, and below which the sliding velocity is treated as static.
The dependency of normalized tangential frictional force, $| \mathbf{F}_{\mathcal{C}}^{t} | / \mu | \mathbf{F}_{\mathcal{C}}^{n} |$, on the relative tangential velocity, $|\mathbf{v}^{t}|$, for different velocity bound, $\epsilon_{v} \in \{0.25, 0.50, 0.75 \}$, is plotted in Fig.~\ref{fig:contactModelPlot}(c).
This numerical treatment can smooth the discontinuity between the static friction and the dynamic friction, such that the elastodynamic system can be solved variationally.

\subsection{Ring-to-rod contact}

The conventional incremental potential formulation could handle the interaction between a ring and a rod if they are both discretized into nodes and edges.
However, the redundant nodal number would slow down the overall computational efficiency.
Thus, a novel barrier functional is introduced here to handle the interaction between a rigid ring and a flexible rod.

\begin{figure}[h!]
  \centering
  \includegraphics[width=0.9\columnwidth]{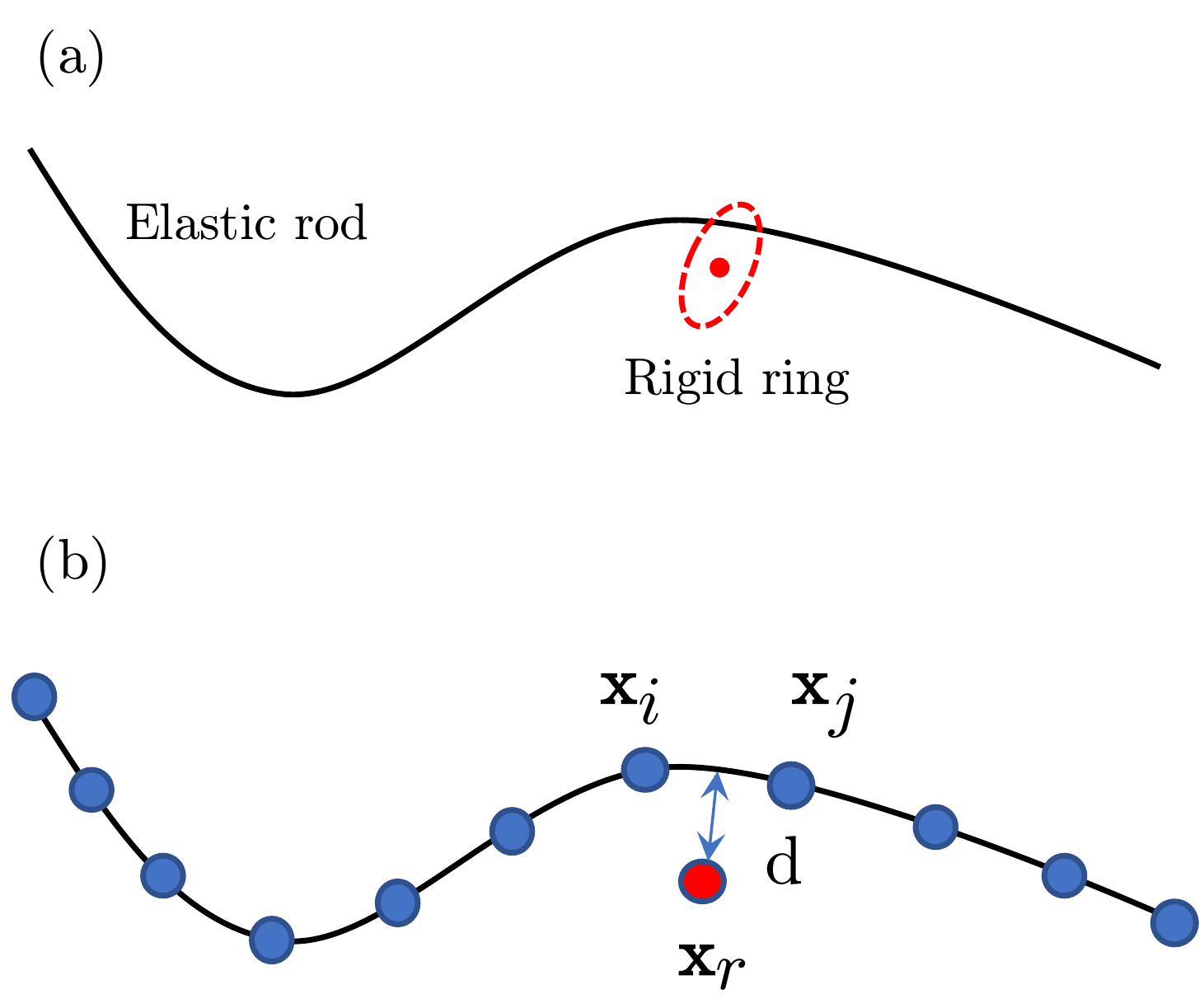}
  \caption{Problem description. (a) A rigid ring slides on a flexible rod. (b) Discrete representation.}
  \label{fig:ringModelPlot}
\end{figure}

As shown in Fig.~\ref{fig:ringModelPlot}, an entangled ring-rod coupled system is considered, and the rod is represented by multiple nodes while the ring is only a single vertex.
The contact element is constructed based on the minimum distance between the sliding ring and the deformable rod, and $3$ nodes are considered here, $\mathcal{J}_{ijr} := \left( \mathbf{x}_{i}, \mathbf{x}_{j}, \mathbf{x}_{r} \right) \in \mathbb{R}^{9}$, where $\mathbf{x}_{i}$ and $\mathbf{x}_{j}$ are the vertices on rod, and $\mathbf{x}_{r}$ represents the position of sliding ring.
The minimum distance $d$ between the joint and the cable here is defined by
\begin{equation}
d = \mathrm{min}\{|\mathbf{x}_{r} - \mathbf{x}_{i}|, |\mathbf{x}_{r} - \mathbf{x}_{j}|, \frac {|(\mathbf{x}_{r} - \mathbf{x}_{i}) \times (\mathbf{x}_{r} - \mathbf{x}_{j})|} {|\mathbf{x}_{j} - \mathbf{x}_{i}|}  \}.
\end{equation}
To ensure the non-deviation condition between the sliding joint and the flexible cable, the following barrier energy functional is constructed by
\begin{equation}
E_{\mathcal{J}} = 
\begin{cases}
K_{\mathcal{J}} \left[ d^2 \log (1 - \frac{d} {\tilde{d}}) \right] & \; \mathrm{ when } \; 0 < d < \tilde{d} \\
+\infty & \; \mathrm{ when } \; d \geqslant \tilde{d}
\end{cases},
\end{equation}
where $K_{\mathcal{J}}$ is the stiffness parameter and $\tilde{d}$ is a barrier parameter similar to the $\hat{d}$ in Eq.(\ref{eq:barrierFunction}).
Correspondingly, the ring radius here is assumed to be zero, and the ring-to-rod distance can be simply modified by minus the ring radius.
It should also be noted that the ring-rod coupled element $\mathcal{J}_{ijr}$ would update with time when the minimum distance between the moving ring and deformable rod is changed.
Again, This newly introduced barrier function for ring-to-rod model is also with $C^{2}$ continuous, and the normal contact force is close to zero when the distance is small, while the force is almost infinite when the distance is close to the threshold, $\tilde{d}$,
\begin{equation}
| \mathbf{F}_{\mathcal{J}}^{n} | = 
\begin{cases}
K_{\mathcal{J}} \left[ 2d \log (1 - \frac{d} {\tilde{d}}) + \frac {d^2}{d - \tilde{d}}  \right] & \; \mathrm{when} \; 0 < d < \tilde{d} \\
+\infty & \; \mathrm{when} \; d \geqslant  \tilde{d}
\end{cases}.
\end{equation}
With this barrier functional, the non-deviation condition between sliding ring joint and elastic cable is successfully achieved.
The dependency of normalized interaction force, $| \mathbf{F}_{\mathcal{J}}^{n} | / K_{\mathcal{J}}$, on the relative distance, $d$, for different barrier parameter, $\tilde{d} \in \{0.50, 0.75, 1.00 \}$, is plotted in Fig.~\ref{fig:contactModelPlot}(b).
The formulation of frictional force for ring-to-rod contact is identical to Eq.(\ref{eq:barrierFriction}).

\subsection{Equations of motion}

This section discusses the time marching scheme for elastodynamic system with non-penetration contact and ring-to-rod interaction.
An elastic structure is discretized into $N$ nodes and treated as a multibody dynamic system, with degrees of freedom $\mathbf{q} \in \mathbb{R}^{3N}$.
Its second order dynamic growing equations of motion is,
\begin{equation}
\mathbb{M} \ddot{\mathbf{q}}(t) + \mathbb{C} \dot{\mathbf{q}}(t) + \mathbb{K} {\mathbf{q}}(t) = \mathbf{F}_{\mathrm{ext}}(t),
\end{equation}
where $\mathbb{M} $ is the time-invariant $3N \times 3N$ diagonal mass matrix, $\mathbb{K} = \nabla^2 E$ (where $E$ is the total elastic potential) is the tangential stiffness matrix, $\mathbb{C} = \alpha \mathbb{M} + \beta \mathbb{K} $ is the Rayleigh damping matrix, and $\mathbf{F}_{\mathrm{ext}}$ is the external force vector (such as the gravity).
Notice that the system is nonlinear, i.e., $\mathbb{C}$ and $\mathbb{K}$ are time variant.

To achieve the non-penetration condition between two approaching nodes as well as the non-deviation condition between sliding joint and elastic rod, the incremental potential and the dissipative friction need to be included.
Geometric detect is updated at each time step to compute $\mathbf{F}_{\mathcal{C}}$ (for particle-particle contact) and $\mathbf{F}_{\mathcal{J}}$ (for ring-to-rod interaction). 
Here, the normal contact force, $\mathbf{F}_{\mathcal{C}}^{n}$ (and $\mathbf{F}_{\mathcal{J}}^{n}$), is treated as a position-dependent elastic force, while the tangential friction force, $\mathbf{F}_{\mathcal{C}}^{t}$ (and $\mathbf{F}_{\mathcal{J}}^{t}$), is considered as a velocity-dependent damping force.

The implicit Euler method is used to numerically update the DOF vector and its velocity from time step $ t_{k} $ to $ t_{k+1} = t_{k} + h$ ($ h $ is the time step size)~\cite{huang2019newmark}:
\begin{subequations}
\begin{align}
{\mathbb{M}} \ddot{\mathbf{q}}(t_{k+1})  &= {\mathbf{F}}_{\textrm{int}}(t_{k+1}) + \mathbf{F}_{\mathcal{C}}(t_{k+1}) + \mathbf{F}_{\mathcal{J}}(t_{k+1}) + \mathbf{F}_d(t_{k+1}) + \mathbf{F}_{{g}}(t_{k+1}) \\ 
{\mathbf{q}}(t_{k+1}) &= {\mathbf{q}}(t_{k}) + h \dot{\mathbf{q}}(t_{k+1}) \\
\dot{\mathbf{q}}(t_{k+1}) &= \dot{\mathbf{q}}(t_{k}) + h \ddot{\mathbf{q}}(t_{k+1})
\end{align}
\label{eq:EulerMethod}
\end{subequations}
where ${\mathbf{F}}_{\textrm{int}} = - \nabla E$ is the internal elastic force, $\mathbf{F}_d = \mathbb{C} \dot{\mathbf{q}}$ is the damping force vector, and $\mathbf{F}_{{g}}$ is the external gravity force vector.

As the interaction potential is based on a $C^2$ continuous formulation, Newton-Raphson is adopted to solve the nonlinear equations of motion, e.g., the Jacobian matrix associated with Eq.(\ref{eq:EulerMethod}) is available.
It is important to note that the damping force and frictional force are treated semi-implicit, i.e., to get the viscoelastic damping force at $t_{k+1}$, the component of Rayleigh damping matrix, $\mathbb{K}$ is evaluated at time $t=t_{k}$, while the velocity, $\dot{\mathbf{q}}$, is evaluated at time $t=t_{k+1}$;
to get the tangential frictional force at $t_{k+1}$, the normal contact, $\mathbf{F}_{\mathcal{C}}^{n}$ (as well as $\mathbf{F}_{\mathcal{J}}^{n}$), is derived at time $t=t_{k}$, while the relative velocity $\mathbf{v}_{t}$ is evaluated at time $t=t_{k+1}$;
all other forces are treated fully implicitly and updated through a classical gradient descent algorithm.

\section{Results}
\label{sec:ResultDemo}

In this section, numerical examples are solved with our newly-introduced discrete model.
The proposed method is implemented with \textit{Eigen} and optimized by compiling with \textit{BLAS} and \textit{LAPACK} backend, and the sparse matrix is solved by \textit{PARDISO} from \textit{Intel's oneAPI Math Kernel Library} (MKL).
A ring slides on a rigid rod is first used to verify our model, and the Young's modulus of 1D rod is reduced to show the nonlinear dynamic interaction between a sliding ring and the deformable structure.
Then, two complex examples are analysed to show the effectiveness of our model for practical engineering applications.
It is worth noting that the sliding ring in the current study is modeled as a single mass point.

\begin{figure}[h!]
  \centering
  \includegraphics[width=0.85\columnwidth]{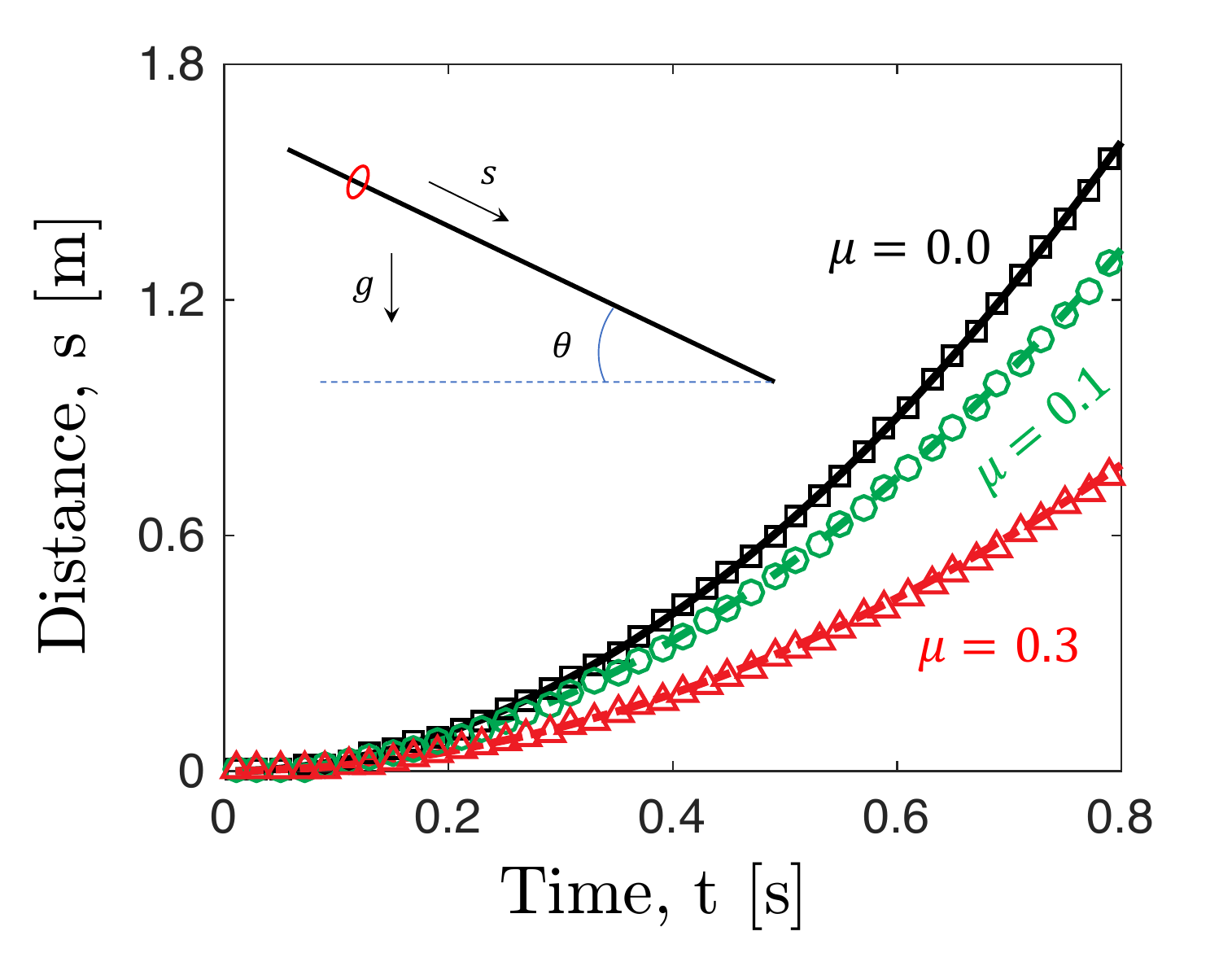}
  \caption{Translational displacements are plotted against time for both numerical results (symbols) and analytical solutions (lines).}
  \label{fig:straightTestPlot}
\end{figure}

\begin{table*}[h!]
\footnotesize
\caption{Physical and geometric parameters for Sec. 3.1, Sec. 3.2, and Sec. 3.3}\label{tableDataSimple}
\begin{tabular}{c|c|c|c|c|c}
Parameters & Notation & Value for Sec. 3.1 & Value for Sec. 3.2 & Value for Sec. 3.3 & Unit \\\hline
Rod modulus & $E$ & $1000$ & $1000$ & $\{100,1,0.01\}$ & GPa \\
Rod length & $L$ & $2.0$ & $0.8$ & $2.0$ & m \\
Rod radius & $r_{0}$ & $10$ & $10$ & $1$ & mm \\
Rod density & $\rho$ & $1000.0$ & $1000.0$ & $1000.0$ & $\mathrm{kg}/\mathrm{m}^3$ \\
Gravity & $g$ & $-10.0$ & $-10.0$ & $-10.0$ & $\mathrm{m}/\mathrm{s}^2$ \\
Ring mass & $m_{0}$ & $1$ & $1$ & $1$ & g \\
Barrier parameter & $\tilde{d}$ & $0.01$ & $0.01$ & $0.01$ & m \\
Stiffness parameter & ${K}_{\mathcal{J}}$ & $1.0$ & $1.0$ & $1.0$ & $\mathrm{N}/\mathrm{m}$ \\
Frictional coefficient & $\mu$ & $\{0.0, 0.1, 0.3 \}$ & $0.0$& $\{0.0, 0.1, 0.3 \}$ & - \\
Velocity bound & $\epsilon_{v}$ & $10^{-6}$ & $10^{-6}$ & $10^{-6}$ & $\mathrm{m}/\mathrm{s}$ \\
Mass damping & $\alpha$ & $0.00$ & $0.00$ & $0.00$ & - \\
Stiffness damping & $\beta$ & $0.00$ & $0.00$ & $0.01$ & -
\end{tabular}
\end{table*}

\begin{figure}[h!]
  \centering
  \includegraphics[width=0.85\columnwidth]{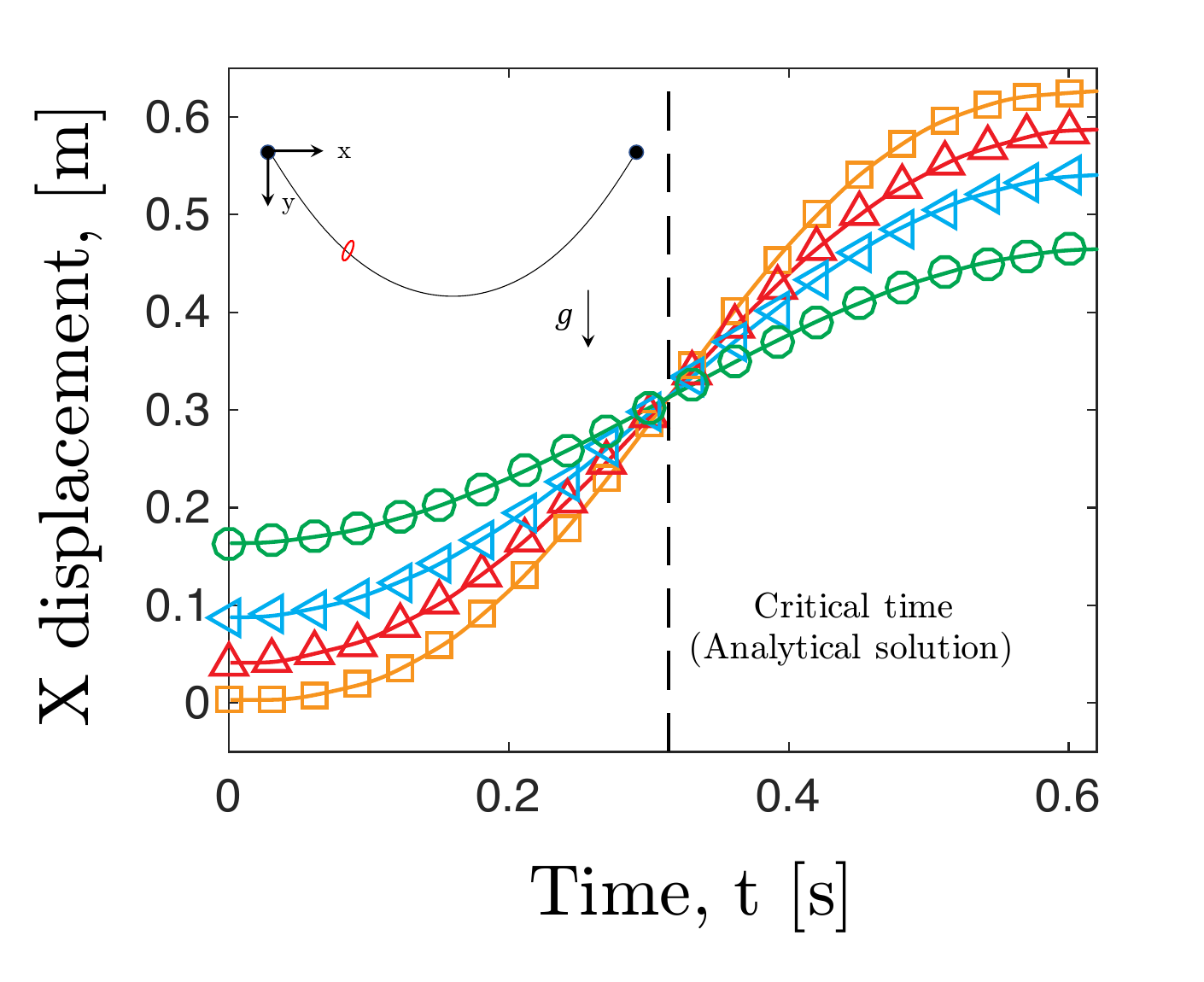}
  \caption{Displacement along $x$ direction is plotted against time for a ring slides on a tautochrone curve.}
  \label{fig:equalTimePlot}
\end{figure}

\subsection{A ring sliding on a rigid inclined bar}
The most basic scenario, which involves a ring sliding on an inclined rigid bar, is examined and compared to an analytical solution.
In this particular case, our focus is solely on the interaction between the ring and the rod, as particle-particle contact does not come into play.
The problem setup is illustrated in Fig.~\ref{fig:straightTestPlot}, and all relevant geometric and physical parameters are detailed in Table~\ref{tableDataSimple}.
The bar is discretized into $200$ nodes, resulting overall DOF vector (rod together with ring) $\mathbf{q} \in \mathbb{R}^{3\times201}$.
The time step size is set to be $h=1$ ms, but a larger time step size is also acceptable.
The inclined angle is selected as $\theta = 30^{\circ}$.
The first node and the last node of the bar are constrained with Dirichlet boundary condition to avoid the rigid body motion.
Due to the relatively large bending stiffness, the rod deformation is negligible and the structure is assumed to be rigid.

The point mass would slide along the tangential direction of the inclined bar if its frictional coefficient is smaller than the threshold, e.g., $\mu < \tan \theta$, and the trajectory of the ring is given by,
\begin{equation}
s = \frac{1} {2} g t^2 (\sin \theta - \mu \cos \theta ).
\end{equation}
Fig.~\ref{fig:straightTestPlot}(b) shows the change of position as a function of time from both numerical data and analytical result.
The agreement of lines (analytical solution) and symbols (numerical data) validates the proposed discrete numerical model.

\begin{figure*}[h!]
  \centering
  \includegraphics[width=0.9\textwidth]{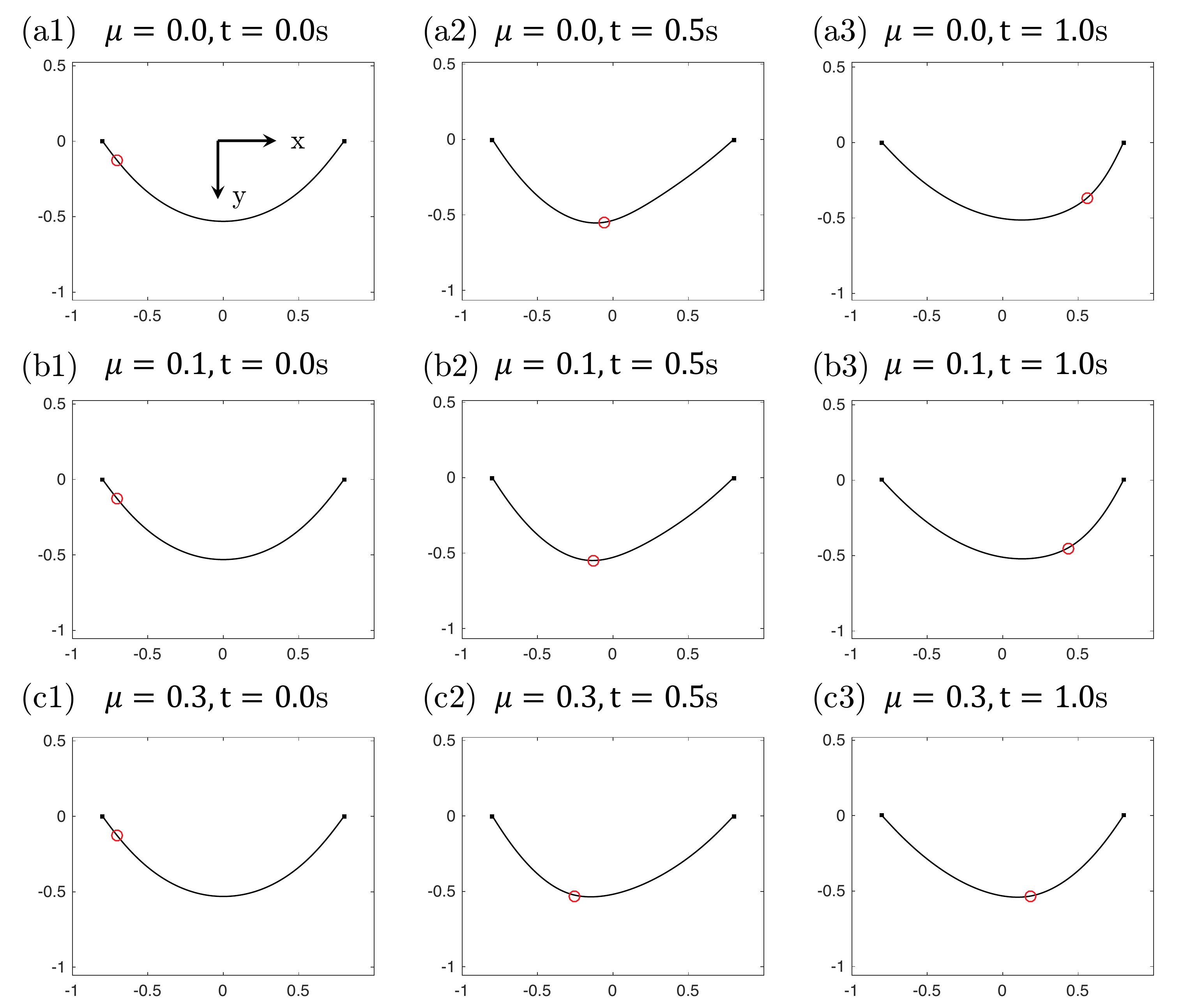}
  \caption{Configurations of a moving mass joint on a flexible rod. Here, the Young's modulus of elastic rod is $E = 1$ GPa.}
  \label{fig:catenaryTestPlot}
\end{figure*}

\begin{figure}[h!]
  \centering
  \includegraphics[width=1.0\columnwidth]{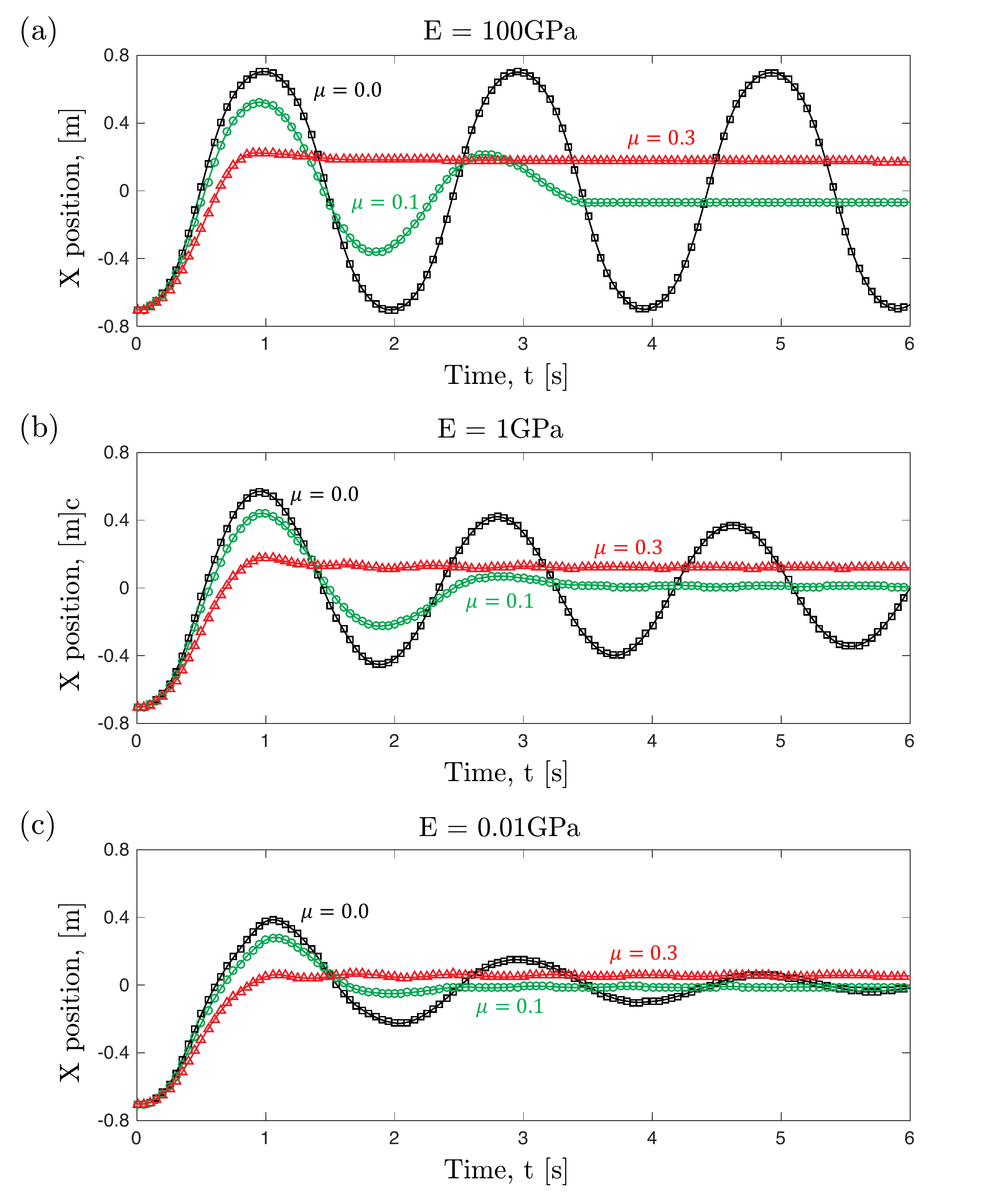}
  \caption{Horizontal position of sliding ring as a function of time. The Young's modulus of elastic rod is varied from $\{100, 1, 0.01\}$GPa and the friction coefficients are selected from $\{0.0, 0.1, 0.3 \}$.}
  \label{fig:catenaryDataPlot}
\end{figure}

\subsection{A ring sliding on a rigid tautochrone curve}
For validation purposes, a more intricate scenario is examined, involving a single ring sliding on a rigid tautochrone curve. It's noteworthy that an analytical solution for this case is also accessible.
The tautochrone curve's shape can be attained by subjecting a flexible rod to compression under the influence of gravity. In its deformable state, it adopts the form of a catenary~\cite{huang2022static}.
Referring to Fig.~\ref{fig:equalTimePlot}, a circular ring entangled with a rigid curve is applied, and the function of which is given by,
\begin{subequations}
\begin{align}
x &= r (\theta - \sin \theta ) \\
y &= r (\cos \theta - 1) \\
\mathrm{with} \; \theta & \in [0, 2\pi].
\end{align}
\end{subequations}
A tautochrone with $r=0.1$ m is adopted, resulting arc-length $L = 0.8$ m.
Again, similar to the previous case, the rod is cut into $200$ nodes and the first and last node of rod are fixed;
The bending stiffness is also enormous to accomplish the structural rigidity.

In Fig.~\ref{fig:equalTimePlot}, we show the $x$ displacement of the ring evolves as a function of time from our numerical side.
Intriguingly, the rings with different start heights, $y$ , reach the minimum with identical time, $T_{\mathrm{critical}} \sim 0.314$ s, as the name of Tautochrone curve.
The analytical solution of drop time can be derived based on classical calculus of variations,
\begin{equation}
T_{\mathrm{critical}} = \pi \sqrt{\frac{r} {g}} .
\end{equation}
A good agreement has been found for the critical time predicted by numerical simulation and analytical formulation. 

\begin{figure*}[h!]
  \centering
  \includegraphics[width=1.0\textwidth]{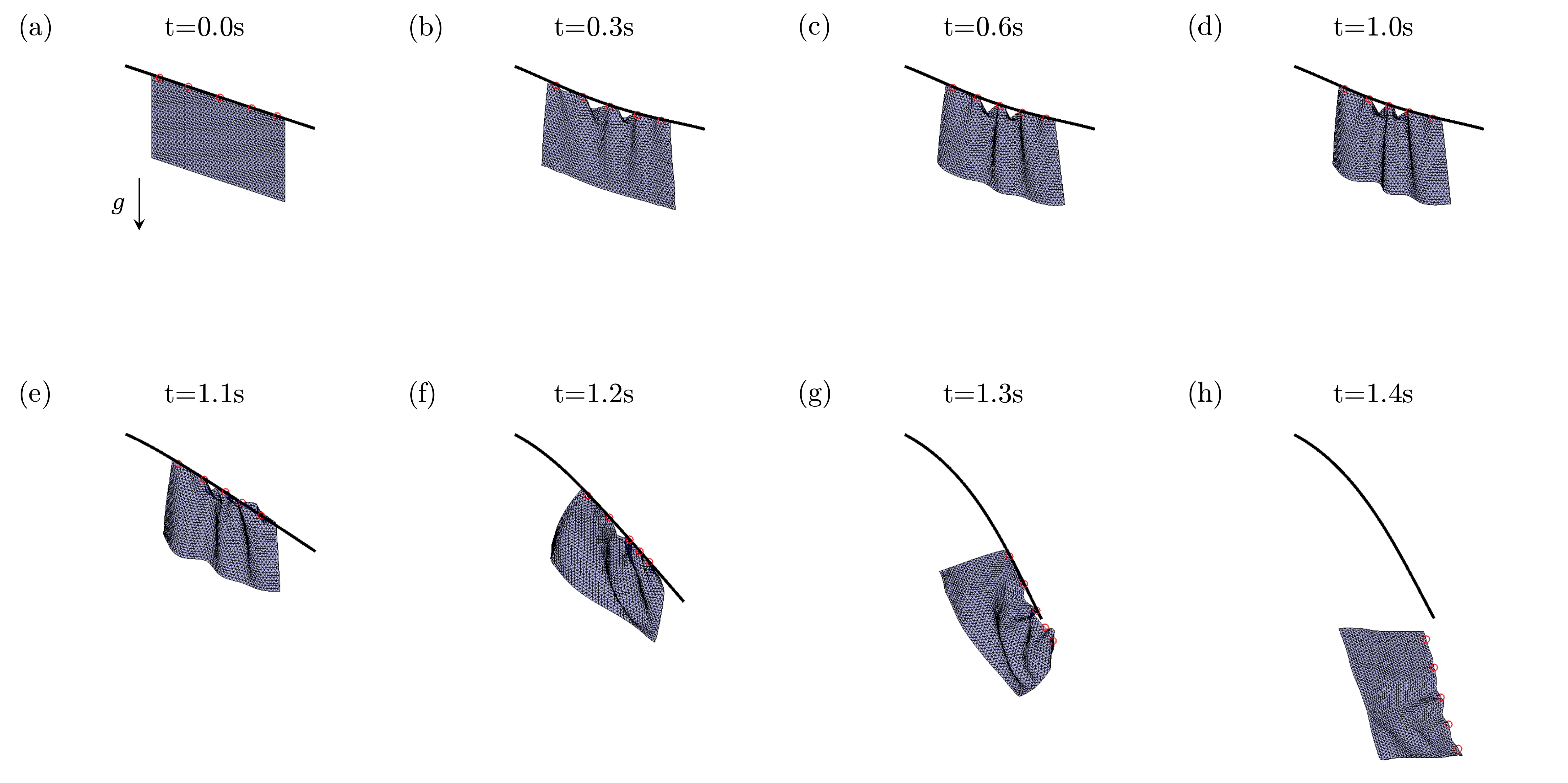}
  \caption{Snapshots of hanging clothes.}
  \label{fig:clothFigurePlot}
\end{figure*}

\subsection{A ring sliding on a flexible catenary curve}

Moving forward, the flexibility of rod and the nonlinear dynamic coupling between ring and rod are illustrated.
Here, a rigid ring moving on a catenary curve is adopted, and the bending stiffness of the suspended flexible rod (which is known as catenary) is significantly reduced such that the structure changes from rigid to elastic.
A circular ring entangled with a catenary curve is applied, and the function of which is given by,
\begin{equation}
y = A \cosh (\frac{x}{A}),
\end{equation}
where its arc-length is $L=2.0$ m.
With a compressive distance ratio $\Delta L / L=20\%$ (and, therefore, $x \in [-0.8,0.8]$ m), the control parameter would be $A = 0.6764$ m.
The geometric and physical parameters are in Table~\ref{tableDataSimple}.
Here, some material damping is added into dynamic system to avoid unrealistic vibratory motion of structure.

\begin{table}[b!]
\footnotesize
\caption{Physical and geometric parameters for Sec. 3.4 }\label{tableDataCloth}
\begin{tabular}{c|c|c|c}
Parameters & Notation & Value & Unit \\\hline
Plate modulus & $Y$ & $0.1$ & MPa \\ 
Rod modulus & $E$ & $1$ & MPa \\
Plate length & $L_{\mathcal{P}}$ & $0.3$ & m \\
Plate width & $W_{\mathcal{P}}$ &  $0.2$ & m \\ 
Plate thickness & $b$ & $1$ & mm \\
Rod length & $L_{\mathcal{R}}$ & $0.5$ & m \\ 
Rod radius & $r_{0}$ & $1$ & mm \\
Material density & $\rho$ & $1000.0$ & $\mathrm{kg}/\mathrm{m}^3$ \\ 
Gravity & $g$ & $-10.0$& $\mathrm{m}/\mathrm{s}^2$ \\
Stiffness parameter & ${K}_{\mathcal{J}}$ & $100.0$ & $\mathrm{N}/\mathrm{m}$ \\
Barrier parameter & $\tilde{d}$ & $0.05$ & m \\
Frictional coefficient & $\mu$ & $0.2$ & - \\
Velocity bound & $\epsilon_{v}$ & $10^{-6}$ & $\mathrm{m}/\mathrm{s}$ \\
Mass damping & $\alpha$ & $0.01$ & - \\
Stiffness damping & $\beta$ & $0.00$ & -
\end{tabular}
\end{table}

The rigid ring is yoked onto the left side of flexible rod and would move to the right due to the existence of gravity, which is similar to the previous scenario.
On the other side, the rod would also be bent as the movement of ring, because the ring can be regarded as a point load onto the curved beam.
Snapshots are provided in Fig.~\ref{fig:catenaryTestPlot}.
In contrast to the previous example, the ring will not move to the same height on the right side if the rod is relatively soft, as some of the gravitational potential is switched to the elastic energy of structure.
Moverover, the energy would dissipate faster and the vibratory motion of ring would gradually disappear as the enlarge of frictional coefficient, as expected.
Seeing Fig.~\ref{fig:catenaryDataPlot} for more details.
 {The convergent study and the performance of the numerical parameters ($\tilde{d}$ and $\epsilon_{v}$) can be found in Appendix A and Appendix B, separately.}

\subsection{Hanging clothes}

Subsequently, the numerical model is employed to tackle a more intricate scenario - the simulation of hanging clothes. This particular scenario has garnered significant attention within the computer graphics community, primarily for generating visually appealing animations~\cite{baraff1998large,choi2005research,bridson2005simulation}.
In this simulation, the 2D surface is divided into multiple triangular meshes, while the 1D rod is discretized into nodes and edges.
All the relevant physical and geometric parameters for this numerical test can be found in Table~\ref{tableDataCloth}.
The elastic rod is represented by $100$ nodes and the flexible cloth is divided into $1402$ nodes (and, therefore, $4090$ connected edges and $2688$ triangular meshes).
The first and last two nodes are manually fixed to establish the clamped-clamped boundary condition; all nodes on cloth are free to move and $5$ massless rings are used to connect thin cloth and slender rod.
It is worth noting that the nodal positions on plate are used directly to represent the sliding rings and no extra vertex is employed to account for coupled joint.
The discrete time step size is $h=0.1$ms.

Initially, the hanging rod would deform from a straight configuration into a curved shape;
next, the sliding joint would move towards to the middle as the rod is inclined and the compressive loads are applied back to cloth;
Finally, some vivid wrinkles would appear due the high flexibility of elastic cloth under compressive load, seeing Fig.~\ref{fig:catenaryDataPlot}(a)-(d) for details.
To show the effectiveness of our ring-to-rod model, the constraints on the right side are released at $t=1.0$ s, such that the clamped-clamped beam would switch to the clamped-free boundary condition (a cantilever beam model).
The hanging cloth would then slide together with the coupled joints from middle to right due to the asymmetric constraints of elastic rod and the external gravitational force;
the rings would gradually shed one by one from the slender rods and loss all connections completely after $t=1.4$ s, referring to Fig.~\ref{fig:catenaryDataPlot}(e)-(h).
The total numerical simulation can be done within $1$ minute.

\begin{figure*}[h!]
  \centering
  \includegraphics[width=1.0\textwidth]{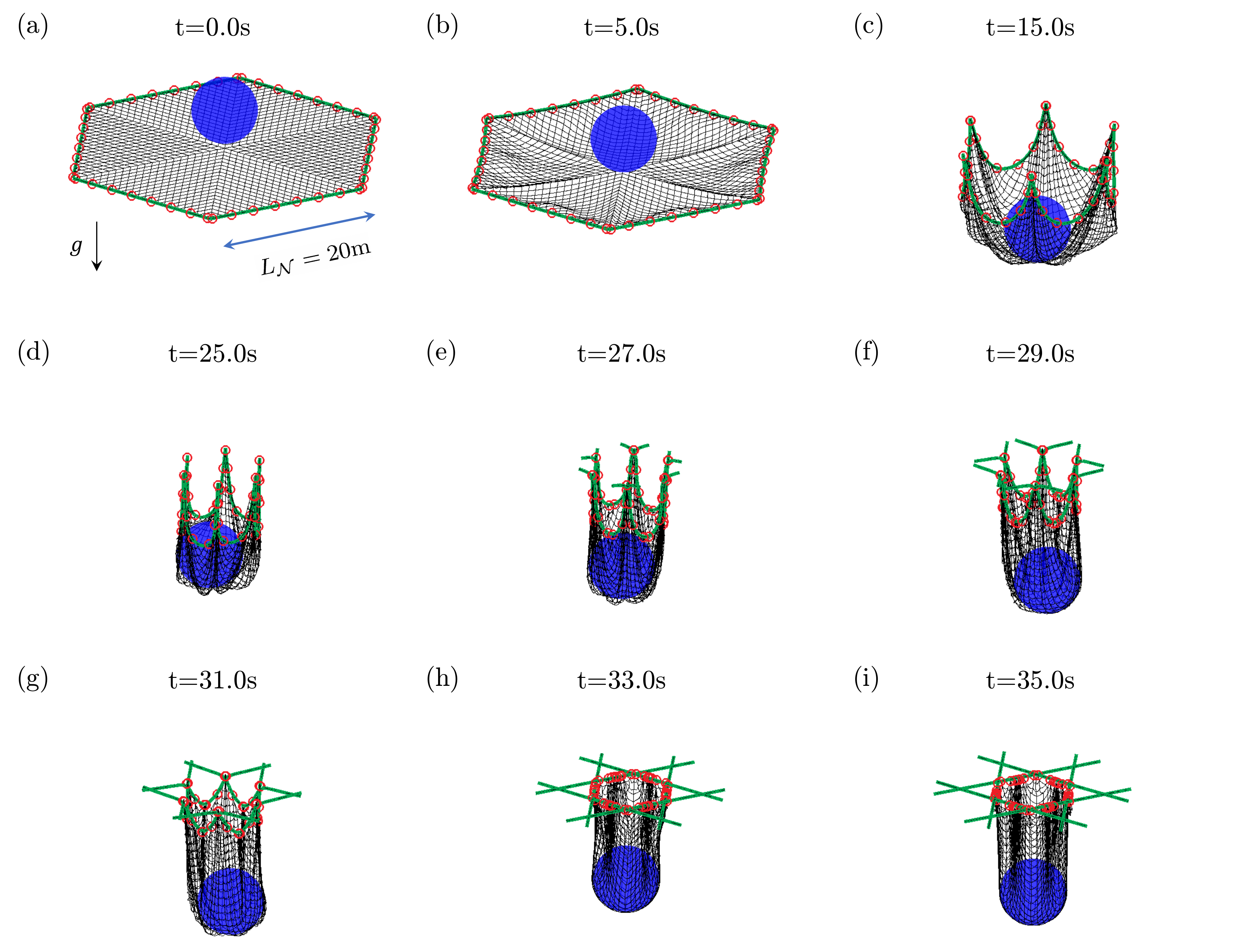}
  \caption{Snapshots of closing a tether-net.}
  \label{fig:netFigurePlot}
\end{figure*}

\subsection{Closing a tether-net}
Lastly, our numerical model is put to use in simulating the dynamic interaction between a tethered rod and a flexible net, involving the use of joint rings. The objective here is to emulate the mechanism by which the net closes during a debris capture mission~\cite{botta2016simulation,sharf2017experiments,hou2021dynamic}.
Referring to Fig.~\ref{fig:netFigurePlot}(a), a rigid object with spherical shape is on top of an elastic net, and the tether rods (coloured by green) for close mechanism are connected with a hexagonal net by multiple rigid rings (coloured by red).
The net is discretized into $6931$ nodes and $8316$ edges;
each tether rod (overall $6$) is divided by $64$ vertex and $63$ edges;
$8$ joint rings are setup for each tether rod.
Again, the ring vertex is the same one with the nodes on net and no extra degrees of freedom are required to mimic the mechanics of rings.
Here, both ring-to-rod interaction and ball-to-net contact are included into the numerical simulation.
The time step size is $h=1$ ms, which is identical to the previous numerical examples.
All other geometric and physical parameters for this numerical experiments are listed in Table~\ref{tableDataNet}.

The overall capture progress can be divided into $3$ phases.
(i) During $0\mathrm{s} \leqslant t < 5\mathrm{s}$, the rigid ball and the flexible net would drop under gravity and contact with each other.
(ii) When $5\mathrm{s} \leqslant t < 25\mathrm{s}$, the six corner nodes are manually moved towards center with speed $1.0\mathrm{m}/\mathrm{s}$, until the distance between adjacent corners is smaller than $5.0$m, e.g., configurations in Fig.~\ref{fig:netFigurePlot}(d).
(iii) After $t=25$s, the six tether rods are stretched out, and, therefore, the net can be pulled up and partially closed due to the connection through joint rings, referring to Fig.~\ref{fig:netFigurePlot}(e)-(i).
The stretching speed of green tether rods is also set to be $1.0\mathrm{m}/\mathrm{s}$ for convenience.
The open size of tether-net system after closing can be tuned during phase (ii).
The total computational time for this case is about $3$ minutes.

\begin{table}
\footnotesize
\caption{Physical and geometric parameters for Sec. 3.5 }\label{tableDataNet}
\begin{tabular}{c|c|c|c}
Parameters & Notation & Value & Unit \\\hline
Net rod modulus & $E_{\mathcal{N}}$ & $1$ & GPa \\
Tether rod modulus & $E_{\mathcal{R}}$ & $10$ & GPa \\ 
Net size & $L_{\mathcal{N}}$ & $20.0$ & m \\
Tether rod length & $L_{\mathcal{R}}$ & $20.0$ & m \\
Rod radius & $r_{0}$ & $1$ & cm \\
Material density & $\rho$ & $1000.0$ & $\mathrm{kg}/\mathrm{m}^3$ \\ 
Ball radius & $R_{\mathcal{B}}$ & $4.0$ & m \\
Ball mass & $M_{\mathcal{B}}$ & $1000$ & kg \\
Ball initial height & $H_{\mathcal{B}}$ & $5.0$ & m \\
Gravity & $g$ & $-10.0$ & $\mathrm{m}/\mathrm{s}^2$ \\
Stiffness parameter & ${K}_{\mathcal{J}}$ & $1000.0$ & $\mathrm{N}/\mathrm{m}$ \\
Barrier parameter & $\tilde{d}$ & $0.5$ & m \\
Stiffness parameter & ${K}_{\mathcal{C}}$ & $10^{5}$ & $\mathrm{N}/\mathrm{m}$ \\
Barrier parameter & $\bar{d}$ & $0.1$ & m \\
Frictional coefficient & $\mu$ & $0.2$ & - \\
Velocity bound & $\epsilon_{v}$ & $10^{-6}$ & $\mathrm{m}/\mathrm{s}$ \\
Mass damping & $\alpha$ & $0.1$ & - \\
Stiffness damping & $\beta$ & $0.0$ & -
\end{tabular}
\end{table}

\section{Conclusion}
\label{sec:Con}

A discrete numerical framework has been created to investigate the nonlinear dynamic interaction that occurs when a mobile ring is placed on an elastic rod.
Rather than subdividing the ring into nodes and edges for the purposes of contact detection and dynamic simulation, a simplified approach was employed, utilizing a single point. This strategy was adopted to achieve model order reduction.
The non-deviation condition between sliding joint and flexible cable was realized by a novel $\log$-type barrier functional.
The inclusion of tangential frictional interaction was subsequently implemented in a manner akin to the treatment of normal contact forces.
The proposed functionals were with higher order continuous and could be adopted to solve the nonlinear elastodynamic system through an implicit approach.
To validate the proposed method, two simple numerical examples were initially solved and their results were compared against analytical solutions.
Then, two examples with large scale systems were considered to demonstrate the effectiveness of the proposed framework for solving practical engineering problems, e.g., cartoon animations in computer graphics and tether-net capture in space debris removal.
The implemented numerical framework for simulating interactions between a ring and a rod has the capability to generate visually impressive videos, catering to the computer graphics community, while also holding potential for enhancing the optimal design of mechanical engineering systems.
In summary, our numerical formulation is well-suited for applications involving the coupling of a ring and a rod system, such as simulating scenarios like hanging cloth and ball capture.
The proposed approach can also be applied to various engineering applications, including configurational force problems, sliding beam configurations, cable-based transportation systems, and space tether systems, among others.
In the future, it is possible to develop a rigid-flexible coupling method by considering the sliding ring as a rigid body, while also incorporating its orientation into the analysis.
%

\begin{acknowledgment}
We are grateful for the financial support from the startup funding from the Newcastle University.
We used the Large Language Model (LLM) – ChatGPT – in the drafting of this manuscript for grammar and language refinement.
\end{acknowledgment}

\section*{Appendix A}

\begin{figure}[h!]
  \centering
  \includegraphics[width=1.0\columnwidth]{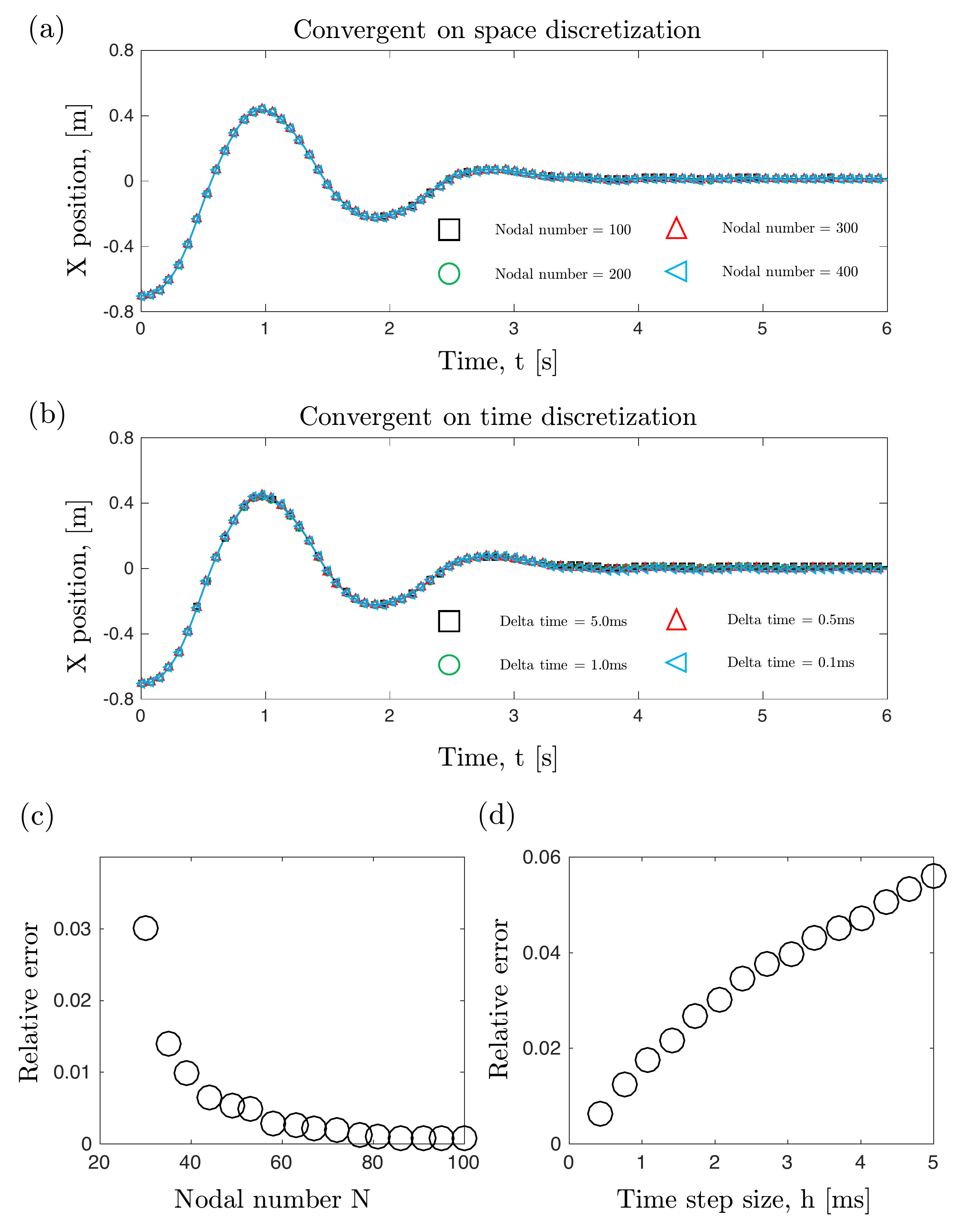}
  \caption{ {Convergent study. (a) Convergent for space discretization. (b) Convergent for time discretization. Relative error as a function of (c) nodal number $N$ and (b) time step size $h$.}}
  \label{fig:convergentPlot}
\end{figure}

{In this appendix, we perform a convergent study in Fig.~\ref{fig:convergentPlot}.
Here, we use $E=1$GPa and $\mu=0.1$ for demonstration (green curve in Fig.~\ref{fig:catenaryDataPlot}(b)).
The overlapping curves in both space and time discretizations indicate the correctness of our model.
Moreover, in Fig.~\ref{fig:catenaryDataPlot}(c) and (d), we show the relative error as a function of nodal number $N$ and time step size $h$, respectively.
The relative error is defined as
\begin{equation}
\mathrm{Error} = \frac { | X(t=1) - X_{\mathrm{real}}(t=1) | } { | X_{\mathrm{real}}(t=1) | },
\end{equation}
where $ X(t=1)$ is the $X$ position of the ring when $t=1$ s and $X_{\mathrm{real}}(t=1)$ is obtained by using a denser mesh ($N=400$) and a smaller time step size ($h=0.1$ ms).
We can clearly see that the relative error can be within the tolerance ($\le 2\%$) when $N>100$ and $h < 1$ ms, and, as this condition, the real-time simulation can be achieved on a desktop processor, i.e., the numerical simulation can be finished within $6$ s for the prediction in Fig.~\ref{fig:catenaryDataPlot} when $N=100$ and $h = 1$ ms.}

\section*{Appendix B}

\begin{figure}[h!]
  \centering
  \includegraphics[width=1.0\columnwidth]{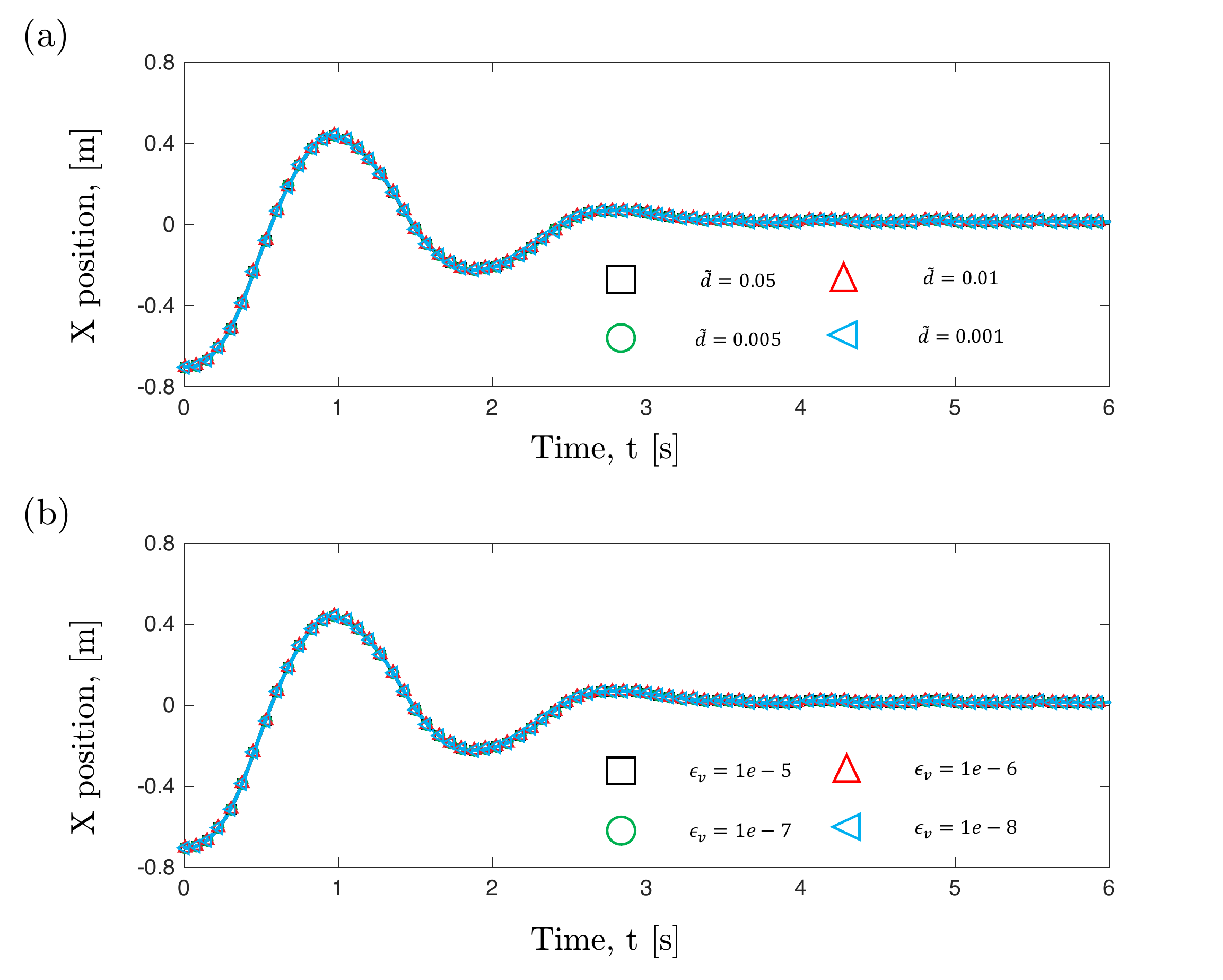}
  \caption{ {Parameter study for (a) $\tilde{d}$ and (b) $\epsilon_{v}$.}}
  \label{fig:convergentParaPlot}
\end{figure}

{In this appendix, we discuss the effect for $\tilde{d}$ and $\epsilon_{v}$.
Again, the green curve in Fig.~\ref{fig:catenaryDataPlot}(b) ($E=1$GPa and $\mu=0.1$) is adopted for demonstration.
On the one side, we vary $\tilde{d} \in \{0.05, 0.01, 0.005, 0.001\}$ m in Fig.~\ref{fig:convergentParaPlot}(a) and find the final result remains unchanged as the decrease of the $\tilde{d}$.
On the other side, a similar approach is used to investigate the influence of $\epsilon_{v}$, and, Fig.~\ref{fig:convergentParaPlot}(b) indicates our numerical prediction can converge as long as a smaller $\epsilon_{v}$ is selected.
In the current study, we use $\tilde{d}=0.01$ m and $\epsilon_{v} = 1e^{-6}$ m/s.}

\bibliographystyle{asmems4}
\bibliography{slideBeam}

\end{document}